\documentclass[letterpaper, 10 pt, conference]{IEEEtran}  % Comment this line out
\usepackage{cite}
                                                         
\IEEEoverridecommandlockouts                         
\usepackage[utf8]{inputenc}
\usepackage[T1]{fontenc}
\usepackage{graphicx}
\usepackage{caption}
\usepackage{subcaption}
\usepackage{amsmath}
\usepackage{amssymb}
\usepackage{ulem}
\usepackage{algorithm}
\usepackage[algo2e]{algorithm2e}
\usepackage{float}
\usepackage{balance}
\usepackage{verbatim}
\usepackage[table,xcdraw]{xcolor}
\usepackage{hyperref}
\usepackage{color}
\usepackage{soul}
\usepackage{algpseudocode}
\usepackage{gensymb}

\makeatletter
\newcommand{\newlineauthors}{%
  \end{@IEEEauthorhalign}\hfill\mbox{}\par
  \mbox{}\hfill\begin{@IEEEauthorhalign}
}
\makeatother

\makeatletter
\def\footnoterule{\kern-3\p@
  \hrule \@width 2in \kern 2.6\p@} % the \hrule is .4pt high
\makeatother

\title{\small 2021 IEEE International Conference on Cloud Computing in Emerging Markets (CCEM)\vspace{15pt}\\
\LARGE \bf
Simulating Realistic MRI variations to Improve Deep Learning model and visual explanations using GradCAM
}

\author{\IEEEauthorblockN{ Muhammed Ilyas Patel*\thanks{*equal contribution from these authors}}
\IEEEauthorblockA{
\textit{IIT Kharagpur}\\
Kharagpur, India\\
jhb.muhammed@iitkgp.ac.in}
\and

\IEEEauthorblockN{\hspace{40pt} Shrey Singla*}
\IEEEauthorblockA{
\hspace{40pt}\textit{IIT Bombay}\\
\hspace{40pt}Mumbai, India \\
\hspace{40pt}190050114@iitb.ac.in}
\and
\IEEEauthorblockN{\hspace{40pt} Razeem Ahmad Ali Mattathodi}
\IEEEauthorblockA{
\textit{\hspace{40pt}IIT Madras}\\
\hspace{40pt}Chennai, India\\
\hspace{40pt}ed17b022@smail.iitm.ac.in}
\and
\IEEEauthorblockN{Sumit Sharma}
\IEEEauthorblockA{\textit{Philips Research} \\
\textit{Philips Healthcare}\\
Bangalore, India \\
sumit.sharma\_1@philips.com}
\and
\IEEEauthorblockN{\hspace{40pt}Deepam Gautam}
\IEEEauthorblockA{\hspace{40pt}\textit{Philips Research} \\
\hspace{40pt}\textit{Philips Healthcare}\\
\hspace{40pt}Bangalore, India \\
\hspace{40pt}deepam.gautam@philips.com}
\and
\IEEEauthorblockN{\hspace{40pt}Srinivasa Rao Kundeti}
\IEEEauthorblockA{\hspace{40pt}\textit{Philips Research} \\
\textit{\hspace{40pt}Philips Healthcare}\\
\hspace{40pt}Bangalore, India \\
\hspace{40pt}srinivasa.rao@philips.com}
}

\begin{document}
\maketitle
\pagestyle{empty}
%%%%%%%%%%%%%%%%%%%%%%%%%%%%%%%%%%%%%%%%%%%%%%%%%%%%%%%%%%%%%%%%%%%%%%%%%%%%%%%%

\begin{abstract}
In the medical field, landmark detection in MRI plays an important role in reducing medical technician efforts in tasks like scan planning, image registration, etc. First, 88 landmarks spread across the brain anatomy in the three respective views- sagittal, coronal, and axial are manually annotated, later guidelines from the expert clinical technicians are taken sub-anatomy-wise, for better localization of the existing landmarks, in order to identify and locate the important atlas landmarks even in oblique scans. To overcome limited data availability, we implement realistic data augmentation to generate synthetic 3D volumetric data. We use a modified HighRes3DNet model for solving brain MRI volumetric landmark detection problem. In order to visually explain our trained model on unseen data, and discern a stronger model from a weaker model, we implement Gradient-weighted Class Activation Mapping (Grad-CAM) which produces a coarse localization map highlighting the regions the model is focusing. Our experiments show that the proposed method shows favorable results, and the overall pipeline can be extended to a variable number of landmarks and other anatomies. 

Keywords - HighRes3DNet, Data Augmentation, Landmark Detection, Medical Image Analysis, Deep Learning, GradCAM

\end{abstract}

%%%%%%%%%%%%%%%%%%%%%%%%%%%%%%%%%%%%%%%%%%%%%%%%%%%%%%%%%%%%%%%%%%%%%%%%%%%%%%%%
\section{INTRODUCTION} 

    \begin{figure*}[h]
        \centering
        \includegraphics[width=0.95\linewidth]{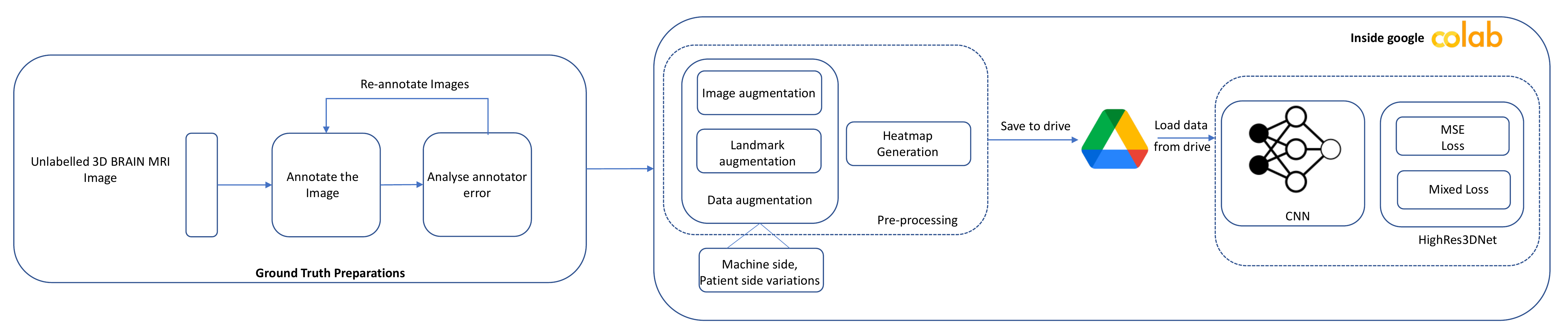}
        \caption{Block diagram shows the pipeline of the proposed work. Ground truth is prepared manually in the first step. Data augmentation, heatmap generation is finished in pre-processing step. Training and result generation is completed in the final step.}
        \label{block}
    \end{figure*}
        Deep neural networks have immensely helped in solving complex real-world problems\cite{liu2017survey}. They have been employed in almost every domain from text recognition\cite{BERT}, speech recognition\cite{deng2013new} to computer vision\cite{voulodimos2018deep} and much more. In medical imaging, conventional image processing pipelines are being replaced by state-of-the-art Deep neural networks\cite{litjens2017survey}. Training medical images require annotated data and high computational resources. The shortage of annotated data in the medical field is a major bottleneck for Deep Learning\cite{singh20203d}. Hence, data augmentation techniques are employed to increase the training dataset size. 
        
        MRI is a popular medical imaging modality. Much attention is put into the MRI workflow to reduce the overall scan time and automate components wherever possible. Being a multi-planar imaging modality, MRI scans can be acquired in any direction by adjusting the magnetic field gradients accordingly. An automatic workflow for scan plane prescription of different landmarks and anatomies is desirable in clinical settings. Thereby reducing MRI exam time and improving image consistency, especially in longitudinal studies. Ideally,
    the plane prescription should be achieved with minimal disruption
    to the existing clinical workflow. Several approaches
    exist in the literature, some relying on fixing an initial localizer scan, either 2-D or 3-D, then finding the correct
    orientation and adjusting the scan planes for future scans
    using image processing methods (conventional or AI). 
    
    Landmark detection is being actively used in various general imaging applications, for example - facial landmark detection for facial analysis tasks, emotion recognition, head pose estimation, etc. In medical imaging, applications include registration and segmentation. Of the many ways to define landmarks, those of interest are:
    i. key point-based (corner, edge, etc.) and ii. Atlas-based/anatomy
    specific. In the anatomical sense, landmarks are defined as points, curves with specific features that are commonly found in every individual with a certain correspondence in location and topology \cite{pose_estimation}. The automatic landmark detection task deals with building algorithms to train models for specific landmarks using annotated data. Then, this trained model is used to find landmarks on unseen 3D brain MRI scans. Recent improvements in deep learning and reinforcement learning motivate to use them in automatic landmark detection. Training
    a landmark detection framework requires an image-ground
    truth pair. In order to create ground truth, landmark annotations are created using manual or semi-automatic methods. Manual annotation is an exhausting task and requires anatomical understanding. Semi-automatic
    approaches might not work for all cases and, at times, require re-adjustment.
    
    This work is an extension to work done in\cite{2020paper}, where
    we have generated ground truth for around 250 images,
    built a new landmark detection model which can detect 88
    landmarks (previously 5) and can be extended to more number
    of landmarks in the future (if needed). Here, we worked on modifying
    the data-augmentation policy for creating machine
    variations and patient side variations which occur in the practical
    sense (realistic data-augmentation), and performed GradCAM
    analysis for a visual explanation of results of our trained deep learning
    model in order to trust the output of the model on un-seen datasets. This work can be easily extended for other anatomies
    (for example-- knee, liver, heart, etc.) and several other imaging
    applications (like registration, etc.)

%%%%%%%%%%%%%%%%%%%%%%%%%%%%%%%%%%%%%%%%%%%%%%%%%%%%%%%%%%%%%%%%%%%%%%%%%%%%%%%%

\section{RELATED WORK}

Natural images taken by the camera are 2-D, and there are several public datasets containing several thousand images that have accelerated the development of the state of the art 2-D networks whether classification\cite{lu2007survey}, segmentation\cite{ashburner2005unified,haralick1985image}, localization, pose estimation \cite{pose_estimation} etc. Medical imaging deals with both 2-D data as well as volumetric data. The problem becomes challenging while dealing with volumetric datasets. Several approaches are used to deal with volumetric 3-D data – 2D model, 2.5 D model, 3-D patch-based approaches\cite{liao2019evaluate}, feeding complete 3-D data. Due to restriction on the GPU memory, it becomes quite challenging to feed the complete volumetric data at a time as the number of parameters increase. HighRes3DNet \cite{highresnet3d} reports state-of-the-art performance for 3-D volumetric brain segmentations. In our work, we have used modified HighRes3DNet architecture for brain landmark localization.

Data augmentation is a powerful technique to improve the robustness of the model. There are several libraries that provide powerful APIs for data augmentation. SimpleITK\cite{lowekamp2013design}, imgaug library\cite{imgaug}, Augmentor\cite{bloice2017augmentor}, albumentations\cite{buslaev2020albumentations}, and Multidim Image Augmentation Framework by Deepmind\footnote{https://github.com/deepmind/multidim-image-augmentation} to name a few. Simple ITK supports spatial as well as non-spatial transformations (Intensity) and requires in-depth knowledge of image processing. Imgaug is a powerful library for data augmentation. It provides support for augmenting landmarks, heatmaps, segmentation maps, bounding boxes, etc., but only supports 2D images.  Albumentations provides complete API for 3D images as well and supports augmentation for keypoints, heatmaps, bounding box, etc.

In our work, we use the TorchIO library\cite{torch_IO} which is specially designed for medical images. It has easy to use API and has transformations covering the machine and patient side variations. This library lacks support for augmenting landmarks; to overcome this limitation; we implement Algorithm \ref{algo}. 

In order to provide a visual explanation of the neural network, Grad-CAM (Gradient-based Class Activation Map) \cite{gradcam} is generally used. Paper\cite{gradcam} presents grad-CAM outputs for the 2-D classification networks. Generally, different layers of the network store different characteristics; lower layers store lines, edges, and simpler features whereas upper layers store high-level features\cite{zeiler2014visualizing}, which seems more natural and has several advantages - makes transfer learning easier by fixing lower layers and modifying the upper layers or modifying the architecture in the final layers\cite{pan2009survey}. Paper \cite{lee2018robust} implements grad-cam on the medical data for tumor localization and compares Pyramidal CAM and Grad-CAM approaches.  In our work, we implement Grad-CAM for our 3-D landmark detection model.
   
%%%%%%%%%%%%%%%%%%%%%%%%%%%%%%%%%%%%%%%%%%%%%%%%%%%%%%%%%%%%%%%%%%%%%%%%%%%%%%%%

\section{MATERIALS AND METHODS}
% change this section
% \hl{This work consists of cloud-based 3D CNN interaction between google colab} \cite{Carneiro2018PerformanceAO} and google drive\cite{drive} for model development and data storage, respectively. It also details the model development inside the Google colab \cite{Carneiro2018PerformanceAO}, shown in Fig.\ref{block}, training the models on different custom loss functions and data augmentation methods and evaluating their performance for landmarks prediction.

The biomedical image data generated at hospitals are usually
uploaded to the cloud, at scheduled times. The proposed approach
at a high level defines a cloud-based architecture for training
and classification. The pipeline used for training
and prediction inside the google colab environment is shown in Fig. \ref{block}. 

\subsection{Cloud Computing} \label{Cloud Computing}

    We ran our experiments on google Colaboratory. The Google Colaboratory infrastructure is hosted on the Google Cloud Platform. Colaboratory notebooks are Jupyter based notebooks and enable users to collaborate on the same notebook. Colaboratory has pre-configured machine learning and deep libraries, relieving the users from tedious setups. We connect the Google Drive storage facility for storage purposes. We augmented our dataset separately and saved it to google drive, and trained our model by feeding the dataset from google drive, storing the model weights and training loss per epoch while training. After a fixed time period, the virtual machine(VM) gets deactivated, and all the runtime configurations and data are lost. So, we introduced checkpoints to overcome this.
    
    Cloud services provide many advantages. However, cloud services entail additional security threats and can have severe consequences if security is breached. Uploading sensitive images to the cloud can pose serious threats to the subject's privacy. Identification of the subject's face using face recognition software is now possible\cite{schwarz2019identification}. The current standards of removing only metadata in medical images may be insufficient to prevent the re-identification of subjects in research. These problems can be addressed using federated learning\cite{ng2021federated}. Implementing a brain landmark detection model in a federated learning paradigm itself is a new research problem and out of scope for this paper.

\subsection{Dataset}\label{dataset}
    In our work, we use the open-source OASIS-3 dataset\cite{websiteq} which contains MRI images of normal aging and Alzheimer’s Disease patients. The original image size is 256 x 256 x 256 voxels, 1 mm slice thickness, and zero spacing between the slices. This dataset is converted to a survey-like image with image size 224 x 224 x 101 voxels with 2.2 mm slice thickness and zero spacing between the slices. For ground truth, we considered 88 landmark points belonging to different brain sub-anatomies. Table \ref{subanatomy table} presents the sub anatomies and their associated landmarks points. We followed the guidelines of a Clinical MR expert and manually annotated 234 Images. We use 30 images for validation, 50 for testing, and the rest are fed into the model after augmentation.

    \begin{table}[]
    \begin{tabular}{|c|c|c|}
    \hline
    \rowcolor[HTML]{C0C0C0} 
    \textbf{Sub anatomy} & \textbf{No. of landmarks} & \textbf{Landmark numbers} \\ \hline
    Frontal Lobe         & 5                         & 1-5                       \\ \hline
    Brain Stem           & 8                         & 6-13                      \\ \hline
    Brain Boundary MSP   & 11                        & 14-24                     \\ \hline
    Corpus Callosum      & 13                        & 25-37                     \\ \hline
    Eye                  & 8                         & 38-45                     \\ \hline
    Brain Axial Boundary & 10                        & 46-55                     \\ \hline
    Temporal Lobe        & 33                        & 56-88                     \\ \hline
    \end{tabular}
    \caption{Table shows distribution of 88 landmarks into 7 brain sub anatomies.}
    \label{subanatomy table}
    \end{table}

\subsection{Ground Truth and inter-annotator agreement} \label{inter annotator agreement}
\begin{figure*}[h]
        \centering
        \begin{subfigure}{0.4\textwidth}
            \includegraphics[width=\linewidth]{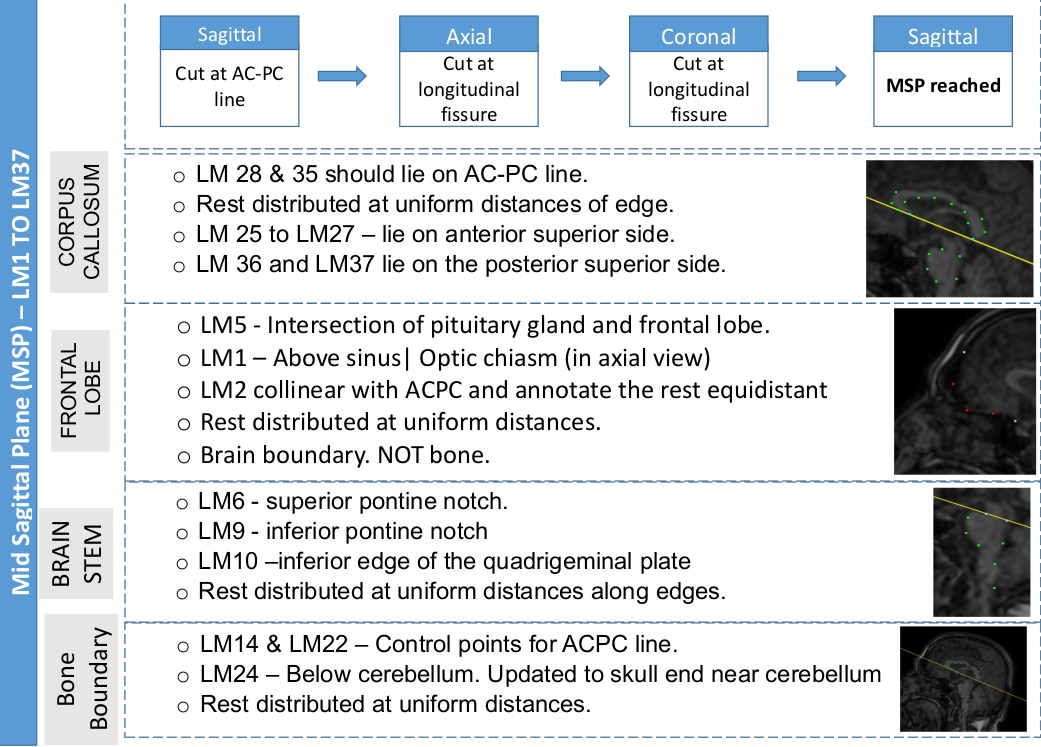}
        \end{subfigure}
        \begin{subfigure}{0.4\textwidth}
            \includegraphics[width=\linewidth]{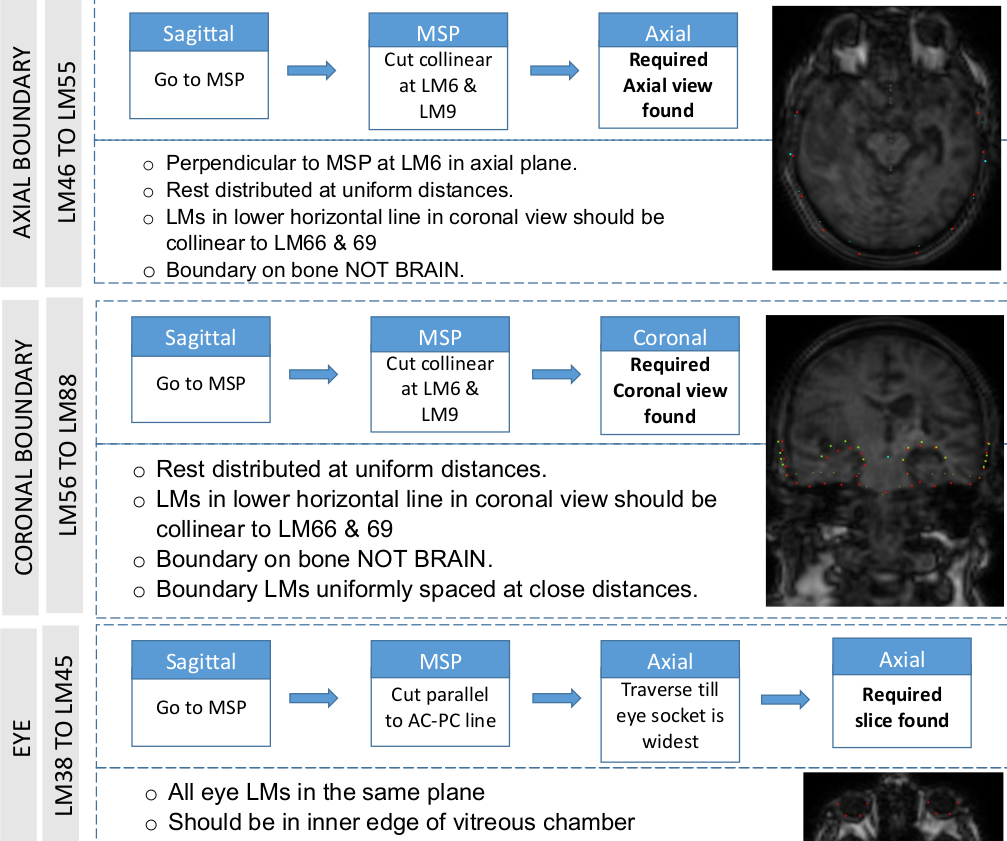}
        \end{subfigure}
        \caption{The figure shows Step-by-step approach followed to annotate MRI data. Sub-anatomy level localization of landmarks ensured minimal inter-annotator variance.}
        \label{annotation guidelines}
    \end{figure*}

    A clinical expert was consulted to get expert input on landmark annotations for various sub anatomies within the brain. The annotations were revised twice. Positioning of specified primary points was given main priority, and secondary points were spread in equal distances along a defined anatomic boundary relative to the control points.
    For ease of annotation, a set of guidelines were laid out for each sub anatomy. The guidelines are more anatomy and image specific rather than landmark specific, and hence only basic anatomical knowledge is a prerequisite for annotation. Fig. \ref{annotation guidelines} gives a detailed overview of the steps followed to annotate the Ground Truth (GT).  
    
    \begin{figure}[h]
        \centering
        \includegraphics[width=.50\textwidth]{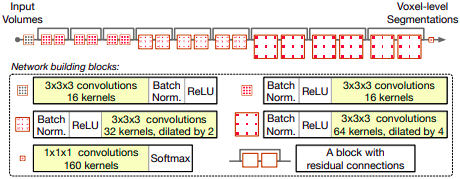}
        \caption{HighRes3DNet architecture with dilated convolutions}
        \label{Model Architecture}
    \end{figure}

\subsection{Data Augmentation} \label{data augmnetation}

    \begin{table}[]
    \begin{tabular}{|l|l|l|}
    \hline
    \rowcolor[HTML]{C0C0C0} 
    \textbf{Tranformation type} & \begin{tabular}[c]{@{}l@{}}Transformation\\ Operations included\end{tabular}                          & variation covered \\ \hline
    Spatial            & \begin{tabular}[c]{@{}l@{}}Elastic Deformation, Affine,\\ Anisotropy\end{tabular}                   & Patient side      \\ \hline
    Non-spatial        & \begin{tabular}[c]{@{}l@{}}Ghosts, Spikes, Bias Field,\\  Noise, Motion artifact, Blur\end{tabular} & Machine side      \\ \hline
    \end{tabular}
    \caption{Table shows transformation operations available in spatial and non-spatial transformations and the variations covered by them.}
    \label{transformation type}
    \end{table}

\begin{figure*}[h]
        \centering
        \includegraphics[width=0.8\textwidth]{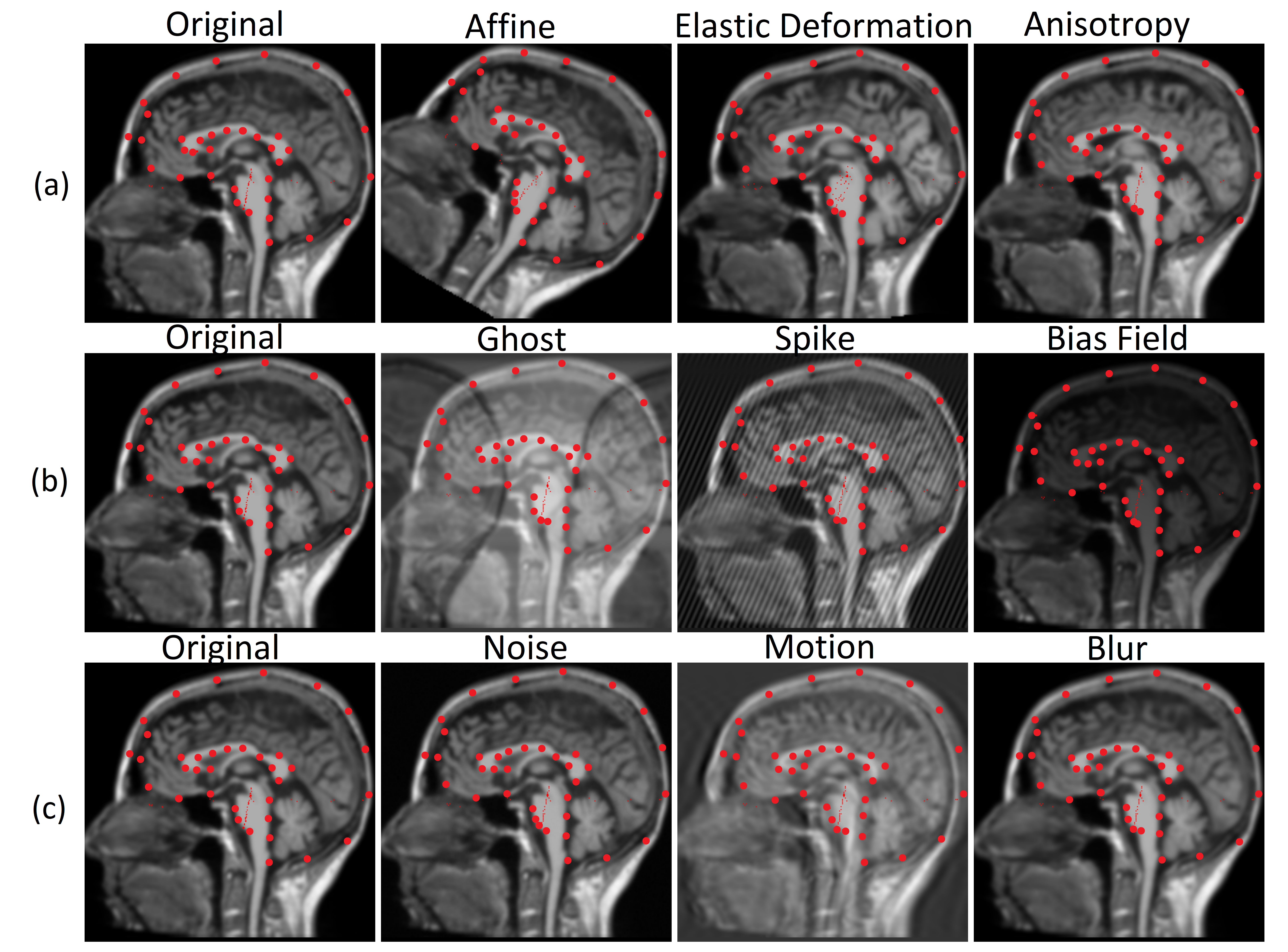} 
        \caption{Row (a) shows patients side variations (spatial transforms), and row (b), (c) shows Machine side variations (non-spatial transforms). The red points (GT) are the landmark points corresponding to different brain sub anatomy.}
        \label{augmentation}%
    \end{figure*}

    Data scarcity is a common problem in Deep Learning. To overcome this, data augmentation is a common practice. It helps to make the model robust and less prone to overfitting. Many times data augmentation technique is more crucial than the model architectures itself\cite{wang2019data}. In general computer vision tasks, common image augmentation techniques include flipping, cropping, scaling, saturation, normalization, etc. However, for medical images, especially MRI, these techniques cannot be used directly. So, while augmenting the medical images, we have to see whether the scenario is possible in the real world. We consulted a clinical applications expert and noted down all the real-life scenarios happening. Mainly there are two kinds of variations. 1. Machine side variations, 2. Patient side variations. Machine side variations include common MRI problems like the occurrence of ghosts\cite{mribook,westbrook2018mri}, spikes, blurs, noise, motion, etc. Patient side variations include anatomical differences, the improper orientation of the patients during the scan, neurodegenerative diseases, etc. Fig. \ref{augmentation} shows various kinds of variations. We use the open-source library TorchIO \cite{torch_IO} to simulate all these variations. \\
    Algorithm \ref{algo} describes the method to augment MRI Images along with the ground truths. We have defined four augmentation policies: DA1, DA2, DA3, DA4. The idea of designing four policies is to cover all possible real-world variations. In real life, MRI scans get affected by machine side and patient side variations. Table \ref{transformation type} shows spatial transformations cover patient side variations and non-spatial transformations cover machine side variations. For an ideal MRI image, the policy DA1 adds one of the patient side variations, and along with that, it adds one of the machine side variations. The policy DA2 only simulates machine-side variations. The policy DA3 simulates patient-side variations. The last policy, DA4, simulates anatomical variations by applying Elastic deformation; thus, more realistic data is generated. Next, we look at how often DA1, DA2, DA3, DA4 occur in real life. Thus, we consulted a clinical expert to give us the probability values to choose a given policy based on what happens in real life. As a result, we chose 0.2, 0.25, 0.25, 0.3 for DA1, DA2, DA3, DA4 respectively. If we have 100 normal images, after applying Algorithm \ref{algo} we get additional 20, 25, 25, 30 augmented images from DA1, DA2, DA3, DA4, respectively. This set of images is well balanced and covers all real-life variations.

    \noindent \textit{\textbf{Augmenting the MRI image along with Ground truth }}: \\ 
    Non-spatial (intensity) transformations do not require any change in the ground truth, but for spatial transformations, as the MRI image transforms (let’s say rotation by 20\degree) we need to transform the ground truth as well (rotation by 20\degree). %All the spatial transformations and non-spatial transformations are stored in a list.%
    For a given image landmarks pair, a policy is chosen randomly as described in Algorithm \ref{algo}. Depending on the policy, transformation functions are applied to the Image as well as the landmarks. For augmenting landmarks, a 3D faux volume is created for each landmark point (by setting the voxel intensity value 1 for the landmark coordinate and the rest of the voxel to 0). Thus, for a given MRI image, there are 88 landmark points and we create 88 faux volumes and apply the same transformation function to all the 88 faux volumes. Finally, from the faux volumes, the landmark coordinates are extracted back. Thus, we obtain transformed landmark points. This process will continue for all the MRI images.

    \SetKwComment{Comment}{/* }{ */}
    
    \begin{algorithm}

    \caption{Data augmentation algorithm} \label{algo}
    \SetKwInOut{Input}{Input}
    \SetKwInOut{Output}{Output}
    \SetKwComment{Commentt}{/* }{ */}
    % caption
    \Input{
    X - Set of 3D MRI Brain Images(224 x 224 x 101)\\
    \hspace{3pt}Y - Set of Landmark points
    }
    \Output{
    X' - Augmented Images \\ 
    \hspace{3pt}Y' - Transformed Landmark Points
    }
    i $\gets$ 0 \\
    \For{Image in X, Landmark in Y}{
        transform = getTransform() \\
        X'[i] $\gets$ transform(Image) \\
        
        \For{$j \gets 1$ to $88$}{
            faux\_volume = create\_faux\_volume(Landmark[k]) \\
            t\_faux\_volume = transform(faux\_volume) \\
            Y'[i][j] = extract\_keypoint(t\_faux\_volume) \\
        }
        
        i $\gets$ i + 1 \\
        
    }

    % \Function{getTransform}{}
    %   transforms = [] 
    %   p = rand() \\
    %   \eIf{$p < 0.2$}{
    %         transforms.append(one\_spatial\_transform)\\
    %         transforms.append(one\_non\_spatial\_transform)
    %   }{
    %       \eIf{$p > 0.2$ and $p < 0.45$}{
    %             transforms.append(one\_non\_spatial\_transform)\\
    %       }{
    %         \eIf{$p > 0.45$ and $p < 0.7$}{
    %             transforms.append(one\_spatial\_transform)
    %         }{
    %             transforms.append(random\_elastic\_deformation)
    %         }
    %       }
    %   }
    % \EndFunction
    
    \end{algorithm}

\subsection{Model architecture and loss function}\label{model}

    The model architecture used here is taken from the HighRes3DNet architecture\cite{highresnet3d}. It uses residual layers and dilated convolutions to capture and improve the receptive field of the model. In addition, it also uses Dropout and Batch normalization layers to handle the problem of overfitting and exploding and vanishing gradients problem.
    The input to the model is an $N\times N\times N$ voxel, and the output is an $K \times N\times N\times N$ tensor, where each $N \times N \times N$ voxel represents one of the $K$ probabilistic heatmaps for each landmark. The loss function is calculated as mean squared error loss between predicted and ground truth coordinates. The spatial softmax function is described in \cite{gradcam} is used to calculate the predicted coordinates. More details on the spatial softmax function can be found in \cite{gradcam}. This loss function does not consider the distance between the heatmaps from which the points were generated. Hence, a different loss function is proposed and implemented, which is a weighted sum of the distance between heatmaps and coordinates extracted from the heatmaps. This loss will be called {\it Mixed Loss}, $L^M$.\\
    Let $\hat{H_i}$ be the predicted heatmap for $i^{th}$ coordinate, $1 \le i \le K$. Let $\hat{p_i}$ be the coordinate extracted from $\hat{H_i}$, ie $\hat{p_i} = spatial\_softmax(\hat{H_i})$. Let $p_i$ be the ground truth coordinates and $H_i$ be the ground truth heatmaps generated from $p_i$. Let $L_c$ be the component of loss due to distance between predicted and ground truth coordinates and $L_h$ be the component of loss due to distance between predicted and ground truth heatmaps. Let $\phi \in \mathbf{R}^{K \times N \times N \times N}$
    $$L_c = \frac{1}{K}\sum_{i=1}^K (p_i - \hat{p_i})^2$$.
    $$\phi_{t,i,j,k} = \frac{e^{\hat{H}_{t,i,j,k}}}{\sum_{i=1}^N \sum_{j=1}^N \sum_{k=1}^N e^{\hat{H}_{t,i,j,k}}}$$
    $$L_h = \frac{1}{K}\sum_{t=1}^K\sum_{i=1}^N\sum_{j=1}^N\sum_{k=1}^N (-H_{t,i,j,k} \times log(\phi_{t,i,j,k}))$$
    $$L^M = \alpha L_h + (1-\alpha) L_c$$
    
    Here, $\alpha$ is a hyperparameter that decides the contribution of heatmap loss in the overall loss function.

\subsection{Gradient Weighted CAM activation}\label{gradcam}
    The GradCAM is a technique to visualize the attention of the model that has been well studied in the domain of classification in 2 Dimensions. It uses gradients of any target concept flowing into the final convolution layer to produce a coarse localization map highlighting the critical regions in the image for predicting the concept. Here, we propose a modification of the GradCAM algorithm for the landmark annotation task in the 3-dimensional domain. It generates the landmark specific high-resolution heatmap volume for an input image volume $I$ which can be used to study the quality of the model and loss functions. It becomes beneficial in the case when the landmarks are close to each other, and we wish to design the models in such a way that the heatmaps generated from GradCAM do not overlap with each other.\\
    Let $H \in \mathrm{R^{K \times N \times N \times N}}$ be the output of the model on $I \in \mathrm{R^{N \times N \times N}}$. Let $A^K \in \mathrm{R}^{T\times N \times N \times N}$ be the activations 
    of $k^{th}$ layer in the model. Let $L^C_{GradCAM} \in \mathrm{R}^{N \times N \times N}$ be the class discriminative localisation map for $C^{th}$ coordinate on input image $I$. Let the output coordinate for $c^{th}$ coordinate be $R_c$.
    $$R_c = argmax_{i,j,k} H_{c,i,j,k}$$
    $$y_c = H_{c,R_c}$$
    $$\alpha_t = \frac{1}{N^3}\sum_{i=1}^N\sum_{j=1}^N\sum_{k=1}^N \frac{\partial y_c}{\partial A^K_{t,i,j,k}}$$
    $$L^C_{GradCAM} = ReLU(\sum_{t=1}^T \alpha_t A^k_t)$$

%%%%%%%%%%%%%%%%%%%%%%%%%%%%%%%%%%%%%%%%%%%%%%%%%%%%%%%%%%%%%%%%%%%%%%%%%%%%%%%%

\section{EXPERIMENTS and RESULTS}

\begin{table*}[h]
\centering
\begin{tabular}{|l|l|l|l|l|}
\hline
\rowcolor[HTML]{C0C0C0} 
\textbf{Configurations}                                                                 & \textbf{\begin{tabular}[c]{@{}l@{}}Before Augmentation \\ + MSE Loss (in mm) \end{tabular}} & \textbf{\begin{tabular}[c]{@{}l@{}}After Augmentation \\ + MSE Loss (in mm)\end{tabular}} & \textbf{\begin{tabular}[c]{@{}l@{}}Before Augmentation \\ + Mixed Loss (in mm)\end{tabular}} & \textbf{\begin{tabular}[c]{@{}l@{}}After Augmentation\\ + Mixed Loss (in mm)\end{tabular}} \\ \hline
\begin{tabular}[c]{@{}l@{}}HighResnet Baseline + \\ uncorrected GT\end{tabular} & 2.32  $\pm$ 1.51 & 1.64 $\pm$ 1.29 & 1.83 $\pm$ 1.4  & 1.71 $\pm$ 1.31 \\ \hline 
\begin{tabular}[c]{@{}l@{}}HighResnet +\\ dropout regularisation + \\ corrected GT\end{tabular} & 1.83 $\pm$ 1.27 & \textbf{1.69 $\pm$ 1.03} & 1.87 $\pm$ 1.34 & \textbf{1.60 $\pm$ 1.21} \\ \hline 
\end{tabular}
\caption{The results on 50 test data showing Mean Absolute Error between predicted and ground truth coordinates}
\label{result}
\end{table*}

\begin{table*}[h]
    \centering
    \begin{tabular}{|l|l|l|}
        \hline
        \rowcolor[HTML]{C0C0C0} 
         \textbf{Subanatomy} & \textbf{\begin{tabular}[c]{@{}l@{}}MAE between ground truth \\and predicted coordinates \\\hspace{35pt}(in mm)\end{tabular}}  &  \textbf{\begin{tabular}[c]{@{}l@{}}RMSE between ground truth\\ and predicted coordinates \\\hspace{35pt}(in mm)\end{tabular}} \\
        \hline
         Frontal lobe & 1.65 $\pm$ 1.09 & 1.95 $\pm$ 1.34 \\
        \hline
         Brain stem & 1.23 $\pm$ 0.78 & 1.46 $\pm$ 0.98 \\
        \hline
         Brain Boundary MSP & \textbf{2.94 $\pm$ 1.98} & \textbf{3.52 $\pm$ 2.43} \\
        \hline
         Corpus Callosum & 1.26 $\pm$ 0.60 & 1.50 $\pm$ 0.73 \\
        \hline
         Eye & 1.23 $\pm$ 0.70 & 1.41 $\pm$ 0.78 \\
        \hline
         Brain Boundary Axial & \textbf{2.15 $\pm$ 1.38} & \textbf{2.51 $\pm$ 1.70} \\
        \hline
         Temporal Lobe & 1.36 $\pm$ 0.68 & 1.58 $\pm$ 0.82 \\
        \hline
    \end{tabular}
    \caption{The table shows anatomy-wise MAE and RMSE between the predicted and the ground-truth coordinates for our best model.}
    \label{tab:Anatomy_error}
\end{table*}

\subsection{Data Augmentation Policy} 
\begin{figure}[h]
        \centering
        \includegraphics[width=.50\textwidth]{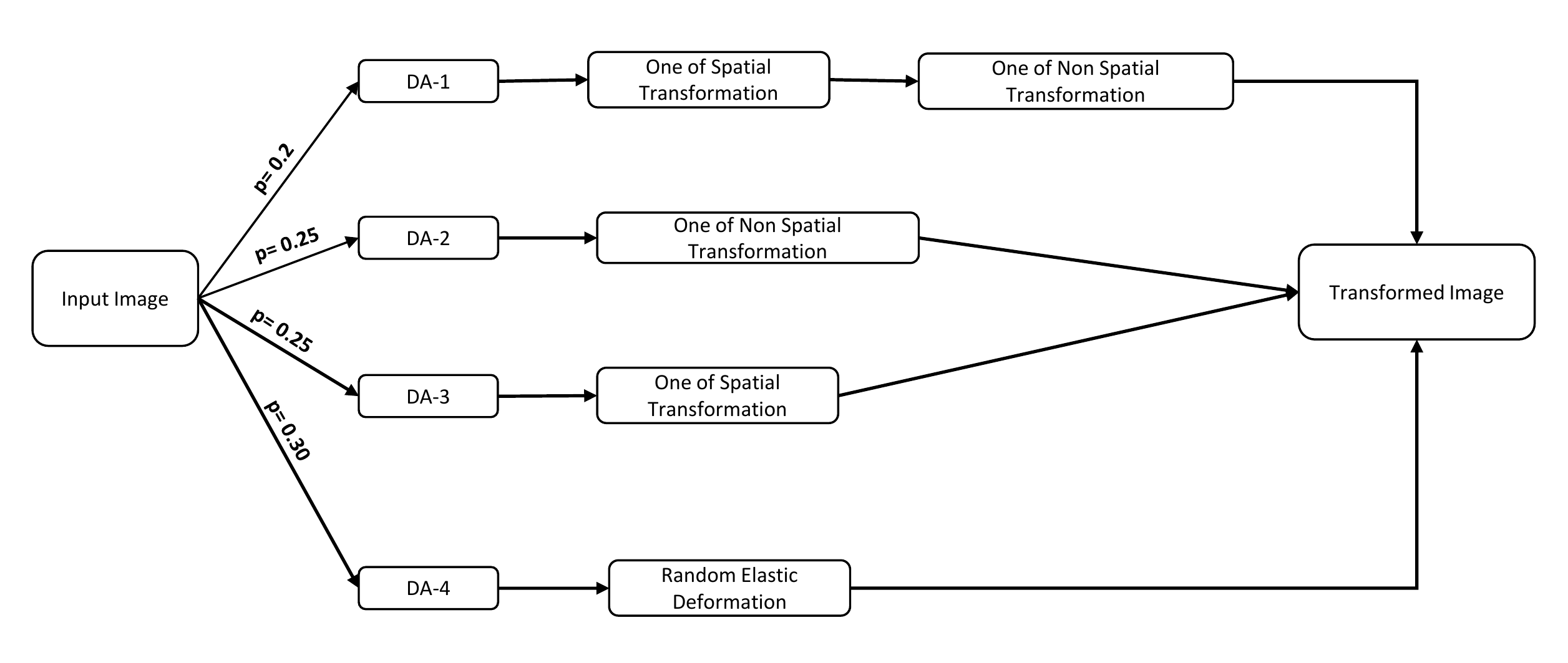}
        \caption{The flowchart shows the data augmentation pipeline. For an input image, we select one out of the four policies at random, based upon the probability (for example, DA1 with p=0.2) as shown, and apply the corresponding transformations to get the transformed image.}
        \label{policy flowchart}
    \end{figure}

    As explained in section \ref{data augmnetation}, we came up with a data augmentation policy to simulate the machine and patient variations. We designed four policies and select a given policy by a given probability value as shown in Fig. \ref{policy flowchart}. 

\subsection{Efforts to resolve the inter-annotator agreement}
\begin{figure}[h]
        \centering
        \begin{subfigure}{0.5\textwidth}
            \centering
            \includegraphics[width=\linewidth]{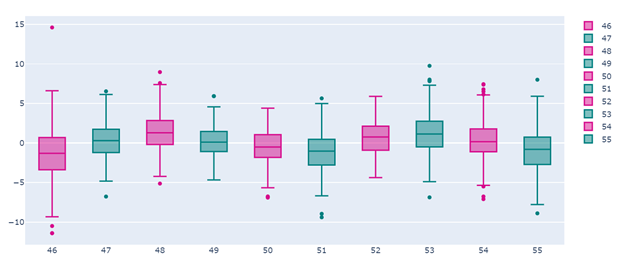}
            \caption{Boxplot shows variations in the ground truth (before correction). All landmarks of the axial boundary (46-55) had high variance due to inter-annotator disagreements.}
            \label{Non corrected Boxplot}
        \end{subfigure}
        
        \begin{subfigure}{0.5\textwidth}
            \centering
            \includegraphics[width=\linewidth]{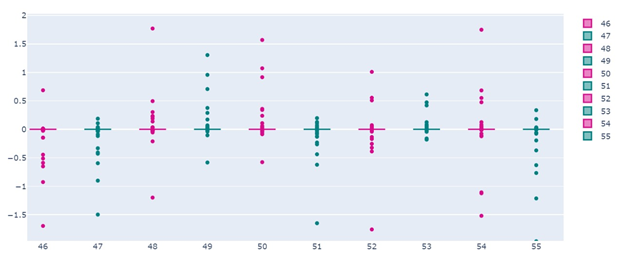}
            \caption{Boxplot shows variations in the ground truth (after correction). All landmarks of the axial boundary (46-55) are confined to a single plane. Hence planarity is ensured.}
            \label{Corrected Boxplot}
        \end{subfigure}
        \caption{Ground truth variations}
    \end{figure}

    After the first set of annotations, the model did not improve notably. An analysis of the annotated data showed disagreements between annotators Fig. \ref{Non corrected Boxplot}. Stricter guidelines were introduced for each sub-anatomy and re-annotated. All 88 landmarks were confined to specific planes for each sub anatomy Fig.\ref{Corrected Boxplot} and boundaries were well defined.
    For example, in the axial view, landmarks 46 to 55 are annotated equidistant to each other and in the same plane around the brain boundary. Initially, the landmarks were not co-planar, as seen in Fig. \ref{Corrected Boxplot}.  Overall, the variance between 88 landmarks was reduced with stricter guidelines. Equidistant is another important factor that needs to be taken care of.  Planarity was the main priority as landmark localization along the same plane for each subanatomy can be transferred to later applications in scan geometry prediction.
    Co-planarity also yielded better results. The model learned on this data showed significant improvement. Hence, it was concluded that inter-annotator agreement and accurate landmark positioning were crucial for better predictions.

\subsection{Results before and after the augmentations}
We use the OASIS-3 dataset for training and evaluation. The model architecture used here is HighRes3DNet with N = 64 and K = 88. The data augmentation distribution is explained in\ref{policy flowchart}. Loss functions used to train the models are MSE Loss and Mixed Loss ($\alpha = 0.4$). We evaluated our model’s performance on the test data of size 50 before and after augmentation. The Table \ref{result} represents the metrics. All the results are presented as Mean Absolute Error between the predicted and ground-truth coordinates. We observe that adding the dropout layer and data augmentation improves the performance of the model significantly. The Mixed Loss function as described above does not show any significant improvement over the MSE Loss function. However, the GradCAM visualization as presented in the next subsection promises better model interpretation on unseen data by training on the Mixed Loss function. We also present the sub-anatomy-wise error in Table \ref{tab:Anatomy_error}. Few sub-anatomies (Brain boundary MSP, Brain Boundary Axial) have a higher error due to difficulty in the annotation. The guidelines of annotations for these sub-anatomies require the annotator to maintain an equal distance between the landmarks and is a difficult task for a human annotator.

\subsection{GradCAM analysis}
    The algorithm to obtain the GradCAM heatmap for an input Image $I$ and a particular landmark $c$ has been described above. We took the HighRes3DNet model architecture and analyzed the quality of heatmaps generated by hooking the activations from the second convolution layer from the third residual block, which happens to be the penultimate layer since it is believed to have the highest level features of the image. The results for each sub-anatomy are presented in the figure. We also compared the qualities of heatmaps generated by two different loss functions - MSE Loss and Mixed Loss, as described above. Comparative study shows that Mixed Loss produces more localized results than MSE Loss and seems to be a better candidate for the loss function.

%%%%%%%%%%%%%%%%%%%%%%%%%%%%%%%%%%%%%%%%%%%%%%%%%%%%%%%%%%%%%%%%%%%%%%%%%%%%%%%%

\section{CONCLUSION AND FUTURE SCOPE}
    \begin{figure*}[h]
        \centering
        \includegraphics[width=0.8\textwidth]{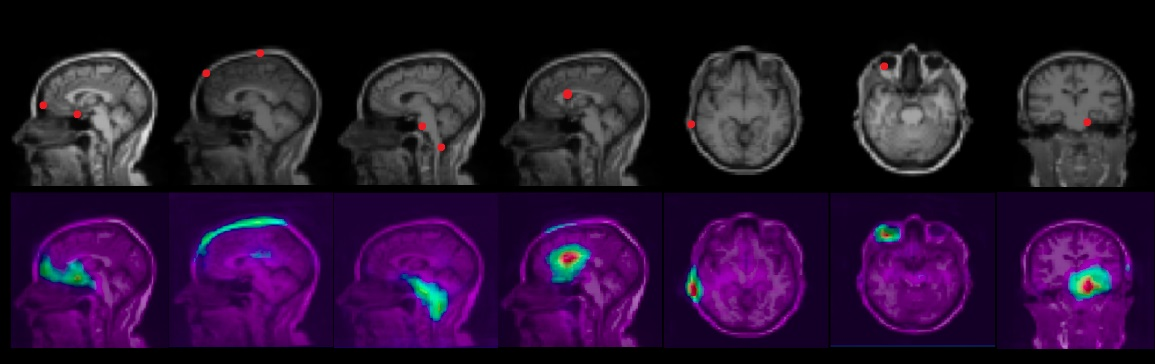} 
        \caption{The top row represents ground-truth landmarks overlaid on the image for each subanatomy, and the bottom row represents the GradCAM heatmaps for a model trained with MSE loss on the same image.}
        \label{Gradcam 1}
    \end{figure*}
    
    \begin{figure*}[h]
        \centering
        \includegraphics[width=0.55\textwidth]{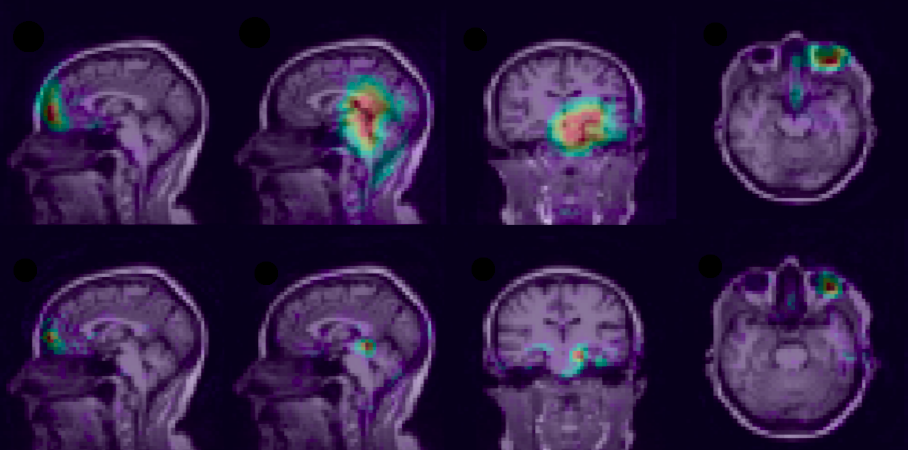} 
        \caption{The figure shows a visual GradCAM heatmap comparison on a test image on four sub-anatomies for our model trained with different loss functions- the top row trained with MSE loss and the bottom row trained with Mixed loss.}
        \label{Gradcam 2}
    \end{figure*}
    
    This paper presents an approach to automatically detect landmarks in medical images and demonstrate the importance of data augmentation. In order to create augmented data, which resembles real MRI artifacts and variations, we formulate and implement our data augmentation policy using TorchIO\cite{torch_IO} library, taking into account both patient side variations and MRI scanner variations. We modify the last layer of HighRes3DNet\cite{highresnet3d} architecture for volumetric landmark detection problem and study the effects of dropout regularisation and custom loss functions on the mean absolute error metrics on test data. The section \ref{dataset} explains the dataset used for experimentation. Section \ref{model}  explains the model architecture and custom loss functions used for training. The section \ref{data augmnetation} gives details about the simulations performed to augment the data. Section \ref{gradcam} summarises the CAM algorithm to generate Gradient weighted CAM activation heatmaps for a particular landmark. We observe that the MAE improves significantly from 2.32 mm to 1.69 mm after implementing data augmentation, dropout regularization and using MSE loss. We also present the GradCAM visualization of the model for all 88 landmarks and verify that the model's focus is as per our expectation for both the models; however, the model trained with Mixed loss focuses more narrowly on the region of interest. Also, different loss functions influence GradCAM results significantly. In our future work, we will focus on improving the model architecture to add skip and attention layers, further improve the loss function, add more variations of different abnormalities in the data augmentation policies, speed up the model training and prediction pipeline. We also plan to introduce federated learning for brain landmark detection problems.
   
\bibliographystyle{IEEEtran}
\bibliography{main}
\end{document}